\documentclass[prl,aps,twocolumn,superscriptaddress,showpacs,floatfix]{revtex4}
\usepackage[dvips]{graphicx}
\usepackage{amsmath}
\usepackage{amssymb}

\newcommand{\kk}{\mathbf{k}}
\newcommand{\pp}{\mathbf{p}}
\newcommand{\qq}{\mathbf{q}}
\newcommand{\KK}{\mathbf{K}}
\newcommand{\PP}{\mathbf{P}}

\newcommand{\PRL}{Phys. Rev. Lett. }
\newcommand{\PRB}{Phys. Rev. B }

\newcommand{\JPSJ}{J. Phys. Soc. Jpn. }

\begin{document}

\title{Exact treatment of exciton-polaron formation by
Diagrammatic Monte Carlo}
\author{Evgeni Burovski}
\affiliation{\mbox{Laboratoire de Physique Th{\'e}orique et
Mod{\`e}les Statistiques, Universit{\'e} Paris-Sud, 91405 Orsay
Cedex, France}} \affiliation{Institut f{\"u}r Physik,
Ernst-Moritz-Arndt-Universit{\"a}t Greifswald, 17489 Greifswald,
Germany}

\author{Holger Fehske}
\affiliation{Institut f{\"u}r Physik,
Ernst-Moritz-Arndt-Universit{\"a}t Greifswald, 17489 Greifswald,
Germany}

\author{Andrei S. Mishchenko}
\affiliation{Cross-Correlated Materials Research Group, RIKEN, 2-1
Hirosawa, Wako, Saitama, 351-0198, Japan}
\affiliation{Russian Research Centre ``Kurchatov Institute'', 
123182 Moscow, Russia}%

\begin{abstract}
We develop an approximation-free Diagrammatic Monte Carlo
technique to study fermionic particles interacting with each other
simultaneously through both an attractive Coulomb potential and
bosonic excitations of the underlying medium. Exemplarily we apply
the method to the long-standing exciton-polaron problem and present
numerically exact results for the wave function, ground-state
energy, binding energy and effective mass of this quasiparticle.
Focusing on the electron-hole pair bound-state formation, we discuss
various limiting cases of a generic exciton-polaron
model. The frequently used instantaneous approximation to the
retarded interaction due to the phonon exchange is found to be of
very limited applicability. For the case of a light electron and
heavy hole the system is well approximated by a particle in the
field of a static attractive impurity.
\end{abstract}


\maketitle


The problem of two quasiparticles (QPs) interacting via
instantaneous Coulomb interaction and a retarded exchange of bosons
is a tremendously difficult stumbling block in solid state
many-particle physics~\cite{Knox,Egri85,Ueta_etal_TheBook}. The
variety of such objects range from exciton-polarons (EX-P) in
semiconductors, where (opposite) charged holes (H) and electrons (E)
couple to lattice vibrations (phonons)~\cite{Egri85}, to more exotic
situations e.g. in the context of high-$T_c$ superconductivity,
where two holes may form a bound state in an 
antiferromagnetically correlated background
due to exchange of spin excitations
(magnons)~\cite{Barentzen}.

First attempts to tackle the EX-P problem are restricted to
low-dimensional cases, and reduce the two QPs to a preformed
structureless QP object~\cite{Ueta_etal_TheBook,MPS_Rashba_Pekar}. Moreover
the rather crude adiabatic approximation was frequently
used~\cite{Ueta_etal_TheBook}, as well as simple variational
approaches~\cite{Sumi}. Quite recently a quantum Monte Carlo study
has been carried out~\cite{Hohenadler2007}, but for a
one-dimensional (1D) model with simplified Coulomb and E/H-phonon
interactions. In any case, a general approximation-free method for
treating a system of interacting QPs in bosonic fields is missing.
Even the interaction of two QPs through an instantaneous potential
creates enormous technical difficulties. For 3D situations with
realistic QP dispersions, at the moment, Diagrammatic Monte Carlo
(DMC)~\cite{exciton2001} and Bethe-Salpeter~\cite{Alb98} methods
seem to be the most promising techniques to address such problems.
Within a Bethe-Salpeter based  approach coupling to phonons can be
introduced only in a less controllable phenomenological
way~\cite{Marini2007} and we are not aware of any generalization to
a true full-scale problem. By contrast, we will demonstrate that a
corresponding generalization of the DMC method can be done in a
rigorous way.

In the present paper we develop such a general approximation-free
DMC technique and apply it for the first time to the highly
non-trivial EX-P problem. Note that different from direct-space
DMC~\cite{Macridin2004} our method is realized
in momentum space and will thus not be restricted to the treatment
of finite systems. Furthermore the proposed momentum space DMC approach
is capable of describing dispersive fermions and bosons, as well as
(long-range) interactions between those objects in any  dimension.
It is used to obtain the EX-P wave function, energy, and
mass, also for the case when an E-H bound state arises due to cooperative
effect of short-range Coulomb attraction and exchange of phonons.
In addition we determine the ground-state phase diagram for
a restricted EX-P model with contact Coulomb and
particle-phonon interaction, discuss important
limiting cases and the validity of approximative solutions.

To this end let us start from the following  Hamiltonian
\begin{eqnarray}
H &=& \sum_\kk  \varepsilon_c(\kk) \, e^{\dagger}_\kk e_\kk + %
    \sum_\kk  \varepsilon_v(\kk)\, h_\kk \, h^{\dagger}_\kk + %
    \sum_\qq \omega_\qq \, b^{\dagger}_\qq b_\qq \nonumber \\
   &&- \sum_{\kk\qq} \left[ \frac{g_e(\qq)}{\sqrt{N}}
   e^\dagger_{\kk-\qq}e_{\kk} +
   \frac{g_h(\qq)}{\sqrt{N}}h^\dagger_{\kk-\qq}h_{\kk}
   \right] \left( b^\dagger_{\qq} + b_{-\qq} \right)\nonumber\\
   &&- \sum_{\pp\kk\kk'} \frac{U(\pp,\kk,\kk')}{N} \,
   e^{\dagger}_{\kk}h^{\dagger}_{\pp-\kk} h_{\pp-\kk'}
   e_{\kk'}
  %
   \;.
\label{Hamiltonian}
\end{eqnarray}
Here $e_\kk$ ($h_\kk$) annihilates an E (H) in the conduction
(valence) band with dispersion $\varepsilon_c(\kk)$
($\varepsilon_v(\kk)$) and $b_\qq$ is the corresponding annihilation
operator for a phonon with momentum $\qq$. In \eqref{Hamiltonian},
$U(\pp,\kk,\kk')$ describes the attractive interband interaction and
$g_e(\qq)$ ($g_h(\qq)$) the E(H)-phonon coupling. $N$ denotes the
number of lattice sites. We work in the thermodynamic limit
$N\to\infty$.

\begin{figure}[tbh]
\includegraphics[width=0.7\columnwidth]{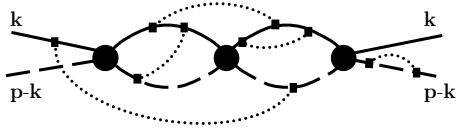}
\caption{A typical diagram for $G^{M=0}_{\pp,\kk}(\tau)$.
Solid (dashed) line represent E (H) propagators, solid circles (squares)
designate Coulomb (QP-phonon) interactions, and dotted lines are
the phonon propagators. Imaginary time runs from left to right.}
\label{fig:diagr}
\end{figure}

An energy ($E_\nu(\pp)$) momentum ($\pp$) eigenstate
($|\nu ; \pp \rangle$) of $H$ can be expressed as linear
combination
\begin{equation}
|\nu ; \pp \rangle = \sum_{M=0}^\infty \sum_{ \{\qq\} \kk } %
\xi^{ M\{\qq \} }_{\pp\kk} (\nu)
{Y^{M\{\qq\}}_{\pp\kk} }^\dagger | \mathrm{vac} \rangle \;
\label{eigenstate}
\end{equation}
of E-H pair basis states in the presence of $M$
phonons (having momenta  $\qq_1,\dots,\qq_M$), where
$Y^{M\{\qq\}}_{\pp\kk} \equiv e_\kk h_{ \pp-\kk-\sum_{j=1}^M \qq_j } 
\prod_{j=1}^{M}
b_{\qq_j}$. The $M=0$ term is understood as
$Y^{M=0}_{\pp\kk}=e_\kk h_{ \pp-\kk}$. Within this basis
we define the two-particle imaginary-time Green function (GF)
with center-of-mass momentum $\pp$ as
\begin{equation}
G^{M\{\qq\}}_{\pp,\kk}(\tau) = \langle \mathrm{vac} |
Y^{M\{\qq\}}_{\pp\kk} (\tau) \, {Y^{M\{\qq\}}_{\pp\kk} }^\dagger(0)
| \mathrm{vac} \rangle \; ,
\label{GF}
\end{equation}
where $Y(\tau)=e^{H\tau}Ye^{-H\tau}$ ($\tau>0$), and
$|\mathrm{vac}\rangle$ is a direct product of the phonon vacuum and
completely filled (empty) valence (conduction) bands. Rewriting
\eqref{GF} in interaction representation and expanding it in terms
of both Coulomb interaction $U$ and E(H)-phonon couplings $g_{e,h}$,
one arrives at a series of phonon-dressed ladder-type Feynman
diagrams~\cite{FetterWalecka}, cf. Fig.~\ref{fig:diagr}. The weight
attributed to a given diagram is the product of the interaction
vertices ($U(\pp,\kk,\kk')$,  $g_e(\qq)$, and $g_h(\qq)$) and
Matsubara GFs of holes, electrons, and phonons with the
corresponding imaginary times and momenta subjected to momentum
conservation imposed by the Hamiltonian (\ref{Hamiltonian}).

In the numerical work, the Monte Carlo updates of diagonal phonon
propagators (connecting two points $\tau_1$ and $\tau_2$ of the same
QP propagator) were performed by the DMC technique developed for a
polaron~\cite{polaron1998,polaron2000}, while Coulomb vertices are
updated as for the pure exciton problem~\cite{exciton2001}. The new
update for nondiagonal phonon propagators (see Fig.~{\ref{fig:UPD})
requires special care to maintain the momentum conservation. When
updating diagonal phonon contributions, momentum conservation is
simply achieved by subtracting the phonon momentum from all QP
propagators between $\tau_1$ and
$\tau_2$~\cite{polaron1998,polaron2000}. Such strategy is not
suitable for the non-diagonal phonon propagators  since the phonon
momentum is taken from one QP line and absorbed by another one. The
problem can be solved, however, by absorbing the momentum transfer
into the Coulomb vertex which, in circular representation for the
GF~\cite{polaron2000,exciton2001}, always appears either to the left
of the phonon propagator ($\tau_a$ in Fig.~\ref{fig:UPD}a) or
between $\tau_1$ and $\tau_2$ ($\tau_b$ in Fig.~\ref{fig:UPD}a). As
illustrated in Fig.~\ref{fig:UPD}b, in the first case the phonon
momentum ${\bf Q}$ is subtracted (added) to (from) the E (H)
propagators located between the Coulomb vertex at $\tau_a$ and
$\tau_1$ ($\tau_2$). In the second case, the E (H) momenta are
changed as $\kk \to \kk -{\bf Q}$ for $\tau' \in [\tau_b, \tau_2]$
($\tau' \in [\tau_1,\tau_b]$), see Fig.~\ref{fig:UPD}c. Note that at
any $\tau'$ the total momentum of E, H, and phonons is equal to the
center-of-mass momentum ${\bf p}$.
\begin{figure}[tbh]
\includegraphics[width=0.6\columnwidth,keepaspectratio=true]{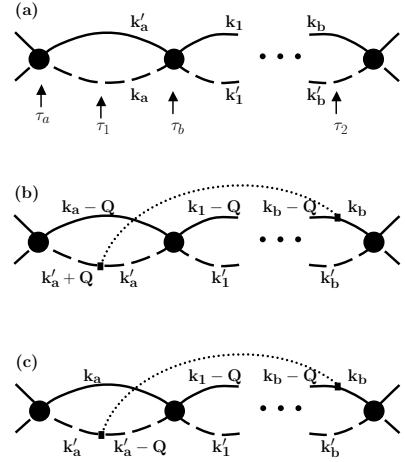}   
\caption{The way how momentum conservation is ensured (b-c) 
when a nondiagonal phonon is added to configuration (a).
Diagrams with a non-diagonal phonon line and no Coulomb vertices
are prohibited by momentum conservation in \eqref{GF}.
}
\label{fig:UPD}
\end{figure}

Now the EX-P's energy, effective mass, and wave function can be
found by DMC sampling of GFs at times 
larger than the reciprocal energy difference between the ground $|
\mathrm{g.s.} \rangle$ and the first excited state,
$\tau>\tau_{\mbox{\scriptsize lim}}$. Inserting the
complete set \eqref{eigenstate} into \eqref{GF}, we have
\begin{equation}
G^{M\{\qq\}}_{\pp,\kk}(\tau) = %
 \sum_{\nu} \left|\xi^{M \{\qq\} }_{\pp\kk}(\nu) \right|^2 e^{-\tau
E_\nu(\pp)} \;.
\label{GF_lehmann}
\end{equation}
For $\tau \ge \tau_{\mbox{\scriptsize lim}}$ the GF projects onto the
ground state in the $\pp$-sector~\cite{exciton2001,polaron1998,polaron2000}:
\begin{equation}
G^{M\{\qq\}}_{\pp,\kk}(\tau \to\infty) \to %
\left|\xi^{M \{\qq\} }_{\pp\kk}(\mathrm{g.s.}) \right|^2 e^{-\tau
E_\mathrm{g.s.}(\pp)} \;.
\label{GF_asympt}
\end{equation}
Due to the normalization $ \sum_{M=0}^\infty \sum_{\{ \qq\} \kk }
\left| \xi^{ M\{\qq \} }_{\pp\kk} (\nu) \right|^2 = 1$, the sum of
all possible $M$-phonon GFs~\cite{polaron2000},
\begin{equation}
\mathfrak{G}_\pp(\tau) = \sum_{M=0}^{\infty} \sum_{\{\qq\}\kk}
G^{M\{\qq\}}_{\pp,\kk} (\tau) \; ,
\label{GF_sum}
\end{equation}
has an especially simple asymptotic form,
$\mathfrak{G}_\pp(\tau\to\infty) \to e^{-\tau
E_\mathrm{g.s.}(\pp)}$. According to \eqref{GF_asympt} and
\eqref{GF_sum} the estimators for the amplitudes $\xi^{ M\{\qq \}
}_{\pp\kk} (\mathrm{g.s.})$ are related to the distribution of
variables $M$, $\{ \qq \}$ and $\kk$ which are all generated by the
DMC algorithm $G^{M\{\qq\}}_{\pp,\kk} (\tau) /
\mathfrak{G}_\pp(\tau) \mid_{\tau \to \infty} \to \left| \xi^{
M\{\qq\} }_{\pp\kk} (\mathrm{g.s.}) \right|^2$. Since the whole set
$\{ \xi^{ M\{\qq \} }_{\pp\kk} (\mathrm{g.s.}) \}$, defining the
inner structure of the EX-P state, is difficult to visualize, we
introduce integrated quantities. For example, the integrated
$Z$-factor $Z^{(M)}_\pp = \sum_{ \kk\{\qq\} } \left| \xi_{\pp\kk}^{M
\{\qq\}} (\mathrm{g.s.}) \right|^2 $ measures the partial weights of
$M$-phonon configurations in the phonon cloud of an EX-P with
center-of-mass momentum $\pp$. On the other hand, for a given
eigenstate $| \nu;\pp \rangle$ of \eqref{Hamiltonian}, the
probability $W(\PP,\KK)$ for a E-H pair to have center-of-mass
momentum $\PP$ and relative momentum $\KK$ is obtained by tracing
out the phonons from the density matrix $\rho = | \nu;\pp \rangle
\langle \nu ;\pp |$:
\begin{equation}
W(\PP,\KK) = \sum_{M=0}^\infty \sum_{ \{ \qq \} } \left|
 \xi^{ M\{ \qq \} }_{\pp\KK } \right|^2
\delta\left( \pp-\sum_{j=1}^{M}\qq_j -\PP
 \right) \; .
\label{p0k0}
\end{equation}

\begin{figure}
\includegraphics[width=0.9\columnwidth,keepaspectratio=true]{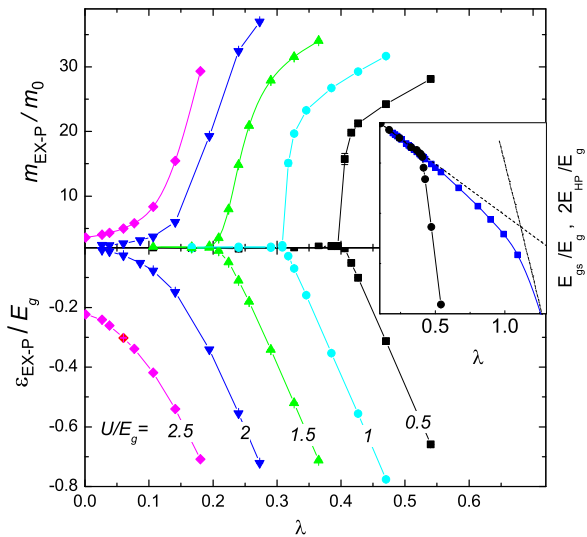}
\caption{(Color online)
Binding energy $\varepsilon_\mathrm{EX-P}$ (in units of the gap
$E_g$) and effective mass $m_\mathrm{EX-P}$ (in units of $m_0 = E_{c,v}/6$)
as functions of E/H-phonon coupling $\lambda$ for $\Omega=0.25 E_g$.
The inset shows
the EX-P energy $E_\mathrm{g.s.}$ (circles) compared to $2E_\mathrm{HP}$
(squares) at $U/E_g=1$.
Statistical error bars are smaller than symbol size.
Dot-dashed lines indicate the HP strong- and weak-coupling results.}
\label{fig:e_and_m}
\end{figure}

In order to validate this novel technique, we now investigate a
minimal 3D simple-cubic (tight-binding) two-band model, $
\varepsilon_{c,v}(\kk) =\widetilde{E}_{c,v} \pm (E_{c,v}/6)
\sum_{\alpha=x,y,z} (1-\cos{k_\alpha}) \; , $ where
$\widetilde{E}_c=E_g$ gives the direct gap at $\kk=0$,
$\widetilde{E}_v=0$, and $E_c$ and $E_v$ are the bandwidths of the
conduction and valence bands, respectively. Furthermore, we take the
phonon frequency $\omega_\qq \equiv \Omega$, particle-phonon
couplings $g_{e,h}(\qq) \equiv g $, and the interband Coulomb
attraction $U(\pp,\kk,\kk')\equiv U$ as momentum independent. A
distinctive feature of this model is that E and H only form a bound
state if $U>U_*$, where $U^{-1}_* = N^{-1}\sum_{\kk \in \mathrm{BZ}}
\left( \varepsilon_c(\kk) - \varepsilon_v({\kk}) - E_g
\right)^{-1}$.
Hence, depending on the Coulomb
attraction $U$ and the dimensionless E/H-phonon coupling $\lambda =
2 g^2 / ( \Omega E_c )$, the  EX-P binding energy
$\varepsilon_\mathrm{EX-P}\equiv E_\mathrm{g.s.} - 2 E_\mathrm{HP}$
(defined as the difference between the EX-P ground-state energy
$E_\mathrm{g.s.}$ and twice the ground-state energy of a single
electron/hole Holstein polaron (HP) $E_\mathrm{HP}$
 at $\pp = 0$ \cite{how_simple}), is either
$\varepsilon_\mathrm{EX-P} =0$ (unbound state) or
$\varepsilon_\mathrm{EX-P} < 0$ (bound state).

Figure~\ref{fig:e_and_m} presents the data for the mass-symmetric
model with $E_c=E_v = 3 E_g$ and $\Omega=0.25 E_g$. Here the
critical Coulomb attraction is $U_*\approx 1.98 E_g$, i.e. for
$U/E_g=2.5$ the H and E are already bound at $\lambda=0$. As
$\lambda$ is increased, the EX-P binding energy and effective mass
smoothly increase and finally show the standard weak- to
strong-coupling crossover~\cite{Hague}. By contrast, for $U < U_*$,
a critical coupling $\lambda_*(U)$ is required to create the EX-P
bound state. While here the single HP  exhibits a rather smooth
crossover from a weakly to strongly mass-renormalized QP at about
$\lambda \approx 1$ (see inset in Fig.~\ref{fig:e_and_m}), the
transition of the E-H pair from unbound to bound state is
accompanied by a much more rapid change of EX-P properties
(Fig.~\ref{fig:e_and_m}). The sharper crossover to the small radius
EX-P regime can be understood from the increasing importance of the
QP-phonon coupling when the EX-QP bound state is
established~\cite{Kishida}. In accordance  with
Ref.~\cite{Macridin2004} we find  $\lambda_*(U=0) \approx 0.5$.

\begin{figure}
 \includegraphics[width=0.99\columnwidth,keepaspectratio=true]{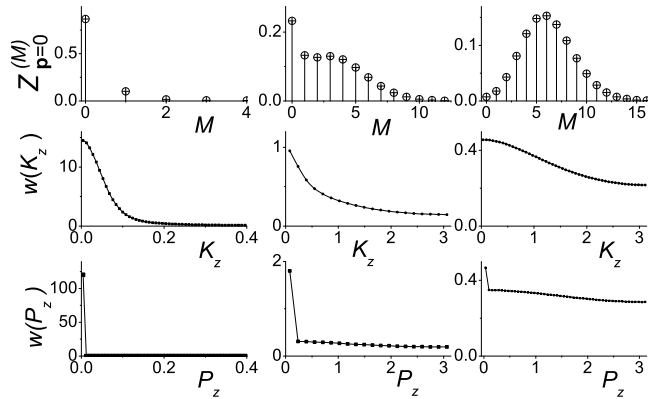}
 \caption{Internal structure of an EX-P
in the mass-symmetric model with  $U=1.5 E_g$
  and
  $\lambda=0.195<\lambda_*$ (1st column),
  $\lambda=0.223 \approx \lambda_*$ (2nd column),
  $\lambda=0.29 > \lambda_*$  (3rd column).
  First row shows the integrated $Z$-factors, second (third) row
  displays the reduced distributions of $W(\PP,\KK)$ \eqref{p0k0}
  $w(K_z) \equiv \int d^3P dK_x dK_y W(\PP,\KK)$
  ($w(P_z) \equiv \int d^3K dP_x dP_y W(\PP,\KK)$).}
  \label{fig:wf}
\end{figure}

The internal structure of the $\pp=0$ EX-P state significantly
changes at the unbound-to-bound-state transition (Fig.\
\ref{fig:wf}). For $\lambda<\lambda_*$ the E-H pair is only weakly
dressed by phonons leading to $\PP \approx 0$. The distribution of
the relative momentum $\KK$ is narrow, pointing to a large real
space separation of the E-H pair. For $\lambda \ge \lambda_*$, E and
H are confined and the $\KK$ distribution broadens. At the same time
the $\PP$ distribution develops a shoulder due to fluctuations of
the momentum of the phonon cloud, and the amplitude of the peak at
$\PP=0$ decreases, reflecting the suppression of the zero-phonon
weight. For $\lambda > \lambda_*$ the phonon distribution $Z^{(M)}$
is almost Gaussian~\cite{polaron2000}. In the $\lambda \approx
\lambda_*$ region, the $Z^{(M)}$-factor distribution shows a kind of
``bimodality'' which, however, is only observed provided that
$\Omega \ll E_{c,v}$. This feature rapidly disappears for larger
phonon frequencies~\cite{MPS_Rashba_Pekar}.

\begin{figure}
\includegraphics[width=0.99\columnwidth,keepaspectratio=true]{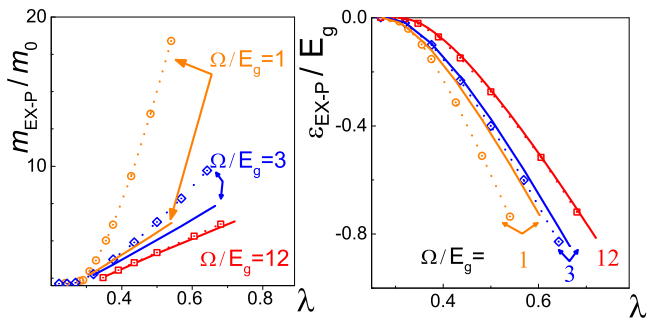}
  \caption{(Color online)
  Exact DMC EX-P energy and effective mass
  (symbols with dotted lines, statistical errors are smaller than the
  symbol size) compared to results for the EAHM
  (solid lines). Data obtained for $U=E_g$ and $E_c=E_v=3E_g$.}
  \label{fig:AHM}
\end{figure}
For finite $\Omega$, the phonon exchange leads to a retarded
interaction between E and H. In the antiadiabatic limit, $\Omega
\geqslant E_{c,v}$, the retardation effects become negligible, and
our model is equivalent to an effective attractive Hubbard model (EAHM) with
$U_\mathrm{eff}=U+2g^2/\Omega$ and hopping inverse proportional to
the single HP mass $t_\mathrm{eff}=1/m_\mathrm{HP}$ (see, e.g.,
Ref.~\cite{Macridin2004}). Comparing the EAHM data with results
obtained for the full model where the retardation effects were included
provides a good check for our DMC algorithm. Indeed we found good
agreement for large phonon frequencies (see Fig.~\ref{fig:AHM} for
$\Omega=12E_g=4E_c$). However, the domain of validity of the
instantaneous approximation is rather limited. As seen from
Fig.~\ref{fig:AHM}, the effective mass, e.g., considerably deviates
from the exact result for $\Omega \simeq E_c$ already.

\begin{figure}
  \includegraphics[width=0.99\columnwidth,keepaspectratio=true]{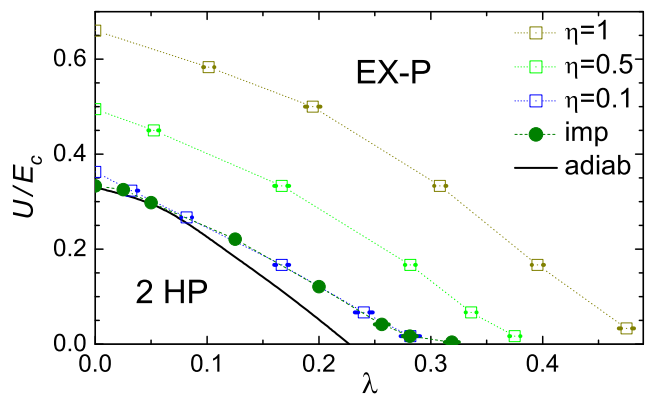}
  \caption{(Color online)
  Phase diagram of the mass-asymmetric EX-P model with
  $E_c=3E_g$, $E_v = \eta E_c$, and $\Omega=E_g$.
  Solid line (solid circles) indicate transition line of
  the static impurity problem with $\eta=0$~\cite{ShinozukaToyozawa1979}
  (obtained by exact numeric techniques~\cite{Holger_Kernel,Mishchenko_new}). Dashed lines are guides to the eye.}
  \label{fig:asym}
\end{figure}

To make contact with the situations for realistic semiconductors we
construct the ground-state phase diagram for a set of mass
asymmetries $\eta= E_v / E_c < 1$ (Fig.~\ref{fig:asym}). The data
clearly show that the larger the bare mass of the hole $m_h^0 \sim
(E_v)^{-1}$, the smaller $U$ and $\lambda$ are required to bind the
excitonic QP. For $\eta \to 0$ ($m_h^0 \to \infty$) the hole becomes
almost immobile and the model is equivalent to that of a
phonon-assisted localization on an attractive impurity
\cite{ShinozukaToyozawa1979}. The excellent agreement of our data
with those obtained for the impurity problem by the Chebyshev space
method~\cite{Holger_Kernel} and DMC in direct space
\cite{Mishchenko_new} provides one more successful test for the
momentum space DMC technique (see Fig.~\ref{fig:asym}).
Surprisingly, the phase diagram of asymmetric EX-P binding well
coincides with that of trapping by impurity already at $\eta=0.1$.

In conclusion, we have developed an exact DMC algorithm for the
interacting electron-hole-phonon system which fully takes into
account an internal structure of the exciton-polaron. The technique
is applicable to any band structure in $D$=1, 2, and 3, an arbitrary
(attractive) Coulomb interaction, and momentum-dependent
fermion-phonon coupling. We constructed the phase diagram for a
phonon-assisted electron-hole bound state formation, and discussed
the change of the internal structure of an exciton-polaron in the
transition regime. Comparison our exact results for the
exciton-polaron problem with findings for an effective attractive
Hubbard model shows that for realistic values of phonon frequencies
the retardation effects can not be neglected.

We appreciate helpful discussions with A.~Alvermann, F.X.~Bronold,
and M. Hohenadler. E.B. acknowledges financial support by DFG
through SFB 652. A.S.M. is supported by RFBR 07-02-00067a.



\begin{thebibliography}{99}
\bibitem{how_simple} The single polaron energy $E_\mathrm{HP}$ is calculated
         by the same computer by code setting $U=0$ and $g_{e}(\qq) =g_{h}(\qq)
\equiv g$.
\end{thebibliography}
\end{document}